\documentclass[fleqn,12pt,twoside]{article}
\usepackage{espcrc1}

\usepackage{graphicx}

\newcommand{\AmS}{{\protect\the\textfont2
  A\kern-.1667em\lower.5ex\hbox{M}\kern-.125emS}}

\title{Collective flow and HBT radii from a full 3D hydrodynamic model
with early chemical freeze out}

\author{Tetsufumi Hirano\address{Physics Department, University of Tokyo, Tokyo 113-0033, JAPAN}
 and Keiichi Tsuda\address{Department of Physics, Waseda University, Tokyo 169-8555, JAPAN}
}
       
\begin{document}

\maketitle

\begin{abstract}
We investigate the effect of early chemical freeze-out on hydrodynamic flow and HBT radii at the RHIC energy with a genuinely three dimensional hydrodynamic model. It is found that the property of early chemical freeze out reduces radial flow, elliptic flow, $R_{\mathrm{out}}$ and $R_{\mathrm{long}}$.
\end{abstract}

\section{INTRODUCTION}

From the particle ratios and the transverse momentum spectra of hadrons in relativistic heavy ion collisions, one can obtain chemical freeze-out temperature $T^{\mathrm{ch}}$ and thermal freeze-out temperature $T^{\mathrm{th}}$ through the analyses of statistical models \cite{QM}.
These temperatures are usually different from each other at the AGS, SPS and RHIC energies, i.e., $T^{\mathrm{ch}} \sim 160$-$200$ MeV while $T^{\mathrm{th}} \sim 100$-$140$ MeV.
This provides the following picture of space-time evolution of hadronic matter: The equilibrated system produced in the early stage of collisions first undergoes the chemical freeze-out where the observed particle ratios are fixed and then goes through the thermal freeze-out where the momentum distribution is fixed.
However, one assumes $T^{\mathrm{ch}}=T^{\mathrm{th}}$ in the conventional hydrodynamic models. As a result, one cannot describe the particle ratios and spectra simultaneously.
As a first trial, we incorporate the early chemical freeze-out into the hadronic equation of state (EOS) \cite{Teaney} for $\mu_{\mathrm{B}}=0$ case \cite{Hirano1} and investigate its effects on hydrodynamic evolution such as radial flow, elliptic flow and HBT radii by using a fully three-dimensional hydrodynamic model (3D hydro) \cite{Hirano2}.

\section{MODEL EOS}

We construct two models EOS to compare the space-time evolution of fluids.
Both models describe the first order phase transition between the QGP phase and the hadron phase at $T_{\mathrm{c}}=170$ MeV.
Here we take the baryonic chemical potential $\mu_{\mathrm{B}} = 0$ for simplicity.
The QGP phase is composed of massless free u, d, s quarks and gluons and is common to both models.
We employ the following two models for the hadron phase.
One is the conventional resonance gas model in which chemical equilibrium is always assumed (the model CE).
We include all baryons and mesons up to the mass of $\Delta(1232)$.
The other model describes the partial chemical equilibrium (the model PCE) in the hadron phase \cite{Bebie}. 
Whether an interaction is chemically equilibrated in the hadron phase depends on its time scale. 
Elastic scattering through a resonance particle, such as $\pi\pi \rightarrow \rho \rightarrow \pi\pi$, $\pi K \rightarrow K^{*} \rightarrow \pi K$ and $\pi N \rightarrow \Delta \rightarrow \pi N$, has large a cross section in the hadronic medium and frequently happens within the lifetime of fluids.
Hence these interactions are equilibrated in the model PCE.
On the other hand, particles which decay through the weak interaction have much longer lifetimes than the typical time scale of the fluid ($\sim 10$ fm/$c$).
So those particles are assumed to be chemically frozen in the model PCE.
Below $T^{\mathrm{ch}}$, we can introduce chemical potential $\mu_i$ associated with the conserved particle numbers. The ratio of the conserved number including contribution from resonance decays $\bar{N}_i (=N_i + \sum_{j\neq i} \tilde{d}_{j\rightarrow i X} N_j)$ to the entropy $S$ has a constant value along an adiabatic path.
Thus we obtain the chemical potential as a function of temperature from the equation $\bar{n}_i(T,\mu_i)/s(T,\{\mu_i\}) = \bar{n}_i(T^{\mathrm{ch}},\mu_i=0)/s(T^{\mathrm{ch}},\{\mu_i\}=0)$,
where $n$ and $s$ are, respectively, the particle number density and the entropy density.
We regard $\pi$, $K$, $\eta$, $N$, $\Lambda$, $\Sigma$ and its anti-particles as ``stable" particles and chemical potentials for all hadrons are represented by chemical potentials associated with these stable hadrons, e.g., $\mu_\rho = 2 \mu_\pi$, $\mu_{K^{*}} = \mu_\pi + \mu_K$, $\mu_\Delta = \mu_\pi + \mu_N$ and so on. We note that the chemical potential of anti-nucleons is equal to that of nucleons in this model: $\mu_{\bar{N}} = \mu_N$.

\section{RESULTS}
We choose initial parameters so as to reproduce the pseudorapidity distribution of charged hadrons for central (0-6\%) and non-central (35-45\%) Au + Au collisions at $\sqrt{s_{NN}}=130$ GeV \cite{PHOBOS1}.
The resultant maximum value of the initial energy density in collisions with vanishing impact parameter is 35 GeV/fm$^3$ at the initial time $\tau_0 = 0.6$ fm/$c$.
For collisions with finite impact parameters, the initial energy density is assumed to scale with the number of binary collisions. For details, see Ref.~\cite{Hirano1}.
\begin{figure}[tb]
\begin{minipage}[t]{75mm}
\includegraphics[width=70mm]{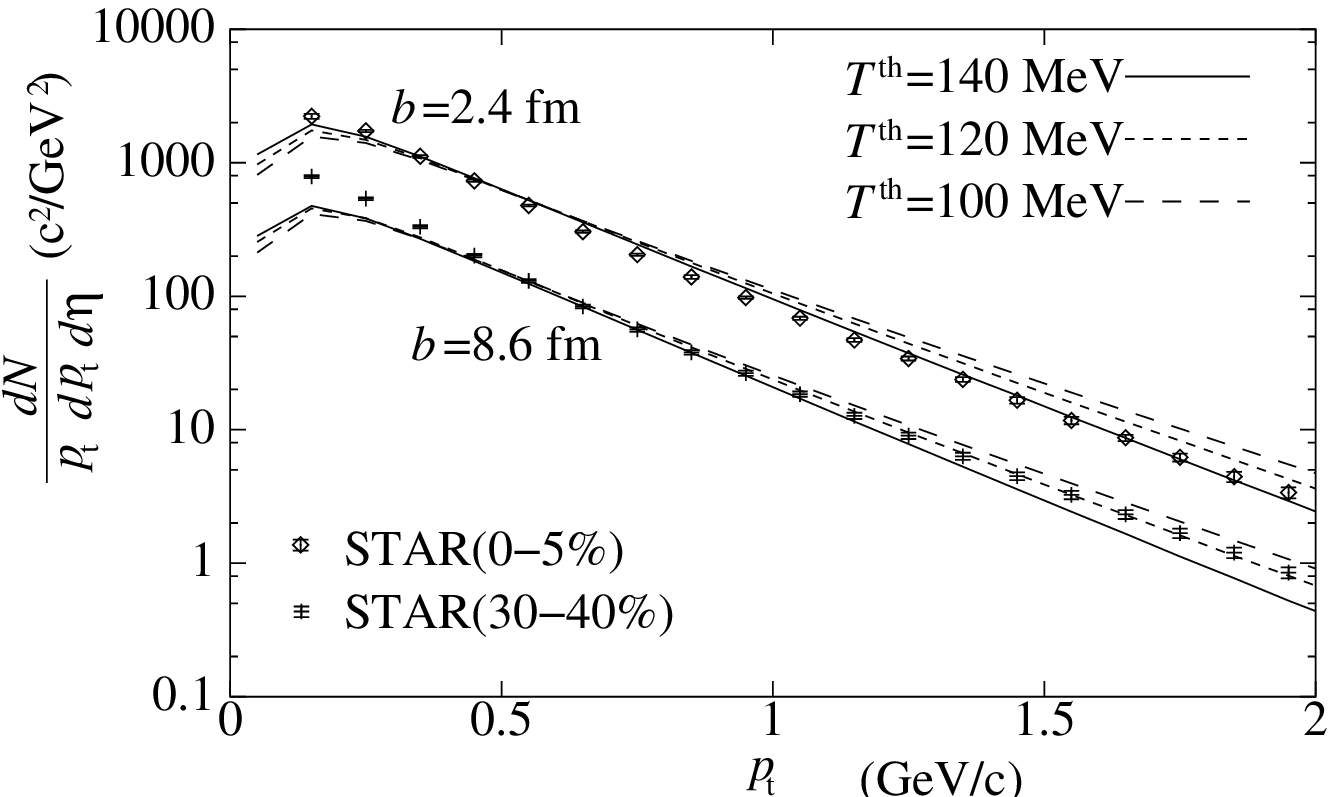}
\end{minipage}
\hspace{\fill}
\begin{minipage}[t]{75mm}
\includegraphics[width=70mm]{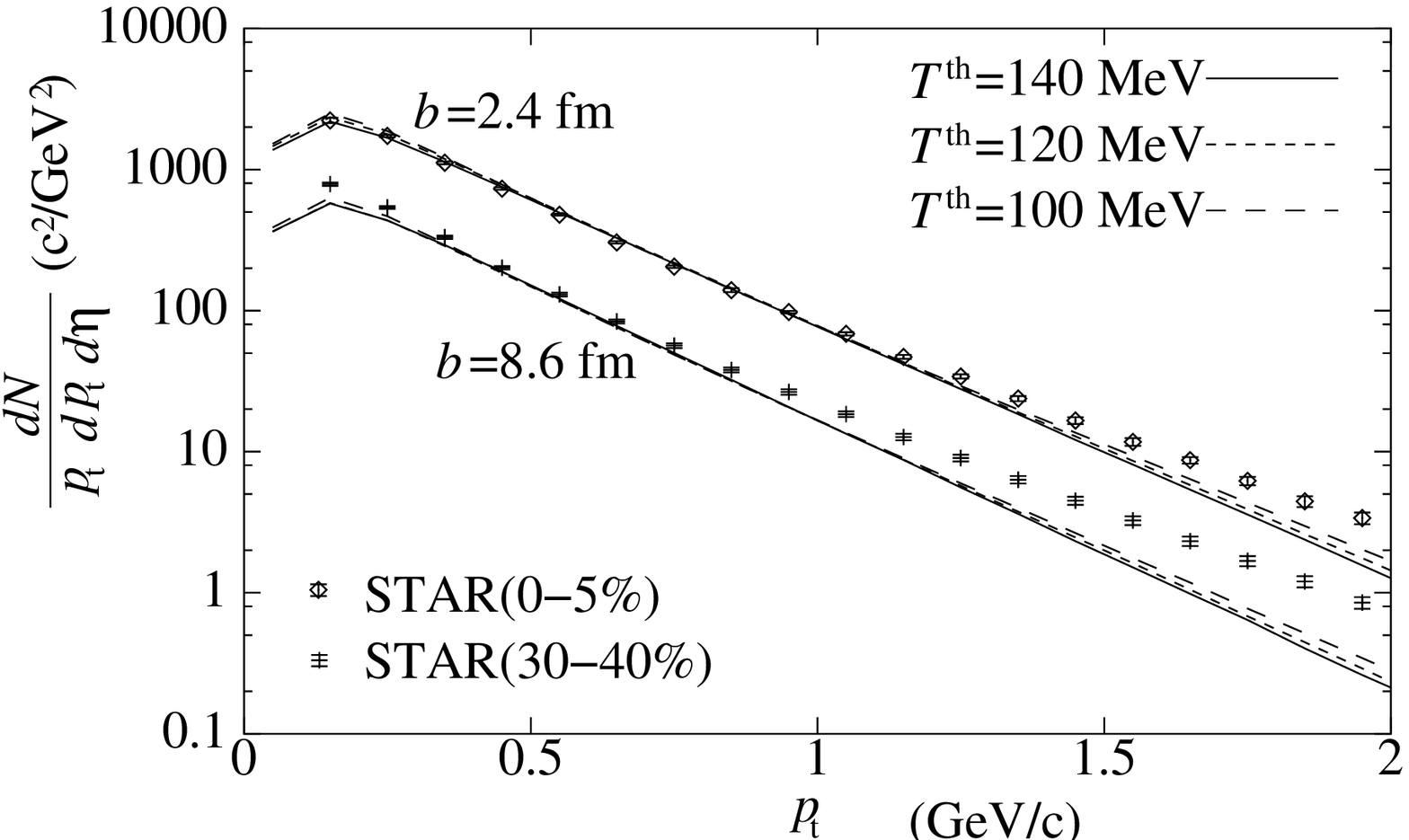}
\end{minipage}
\caption{Transverse momentum distributions of negative charged hadrons in Au + Au central collisions at $\sqrt{s_{NN}}=130$ GeV for the model CE (left) and for the model PCE (right). Solid, dotted and dashed lines correspond to $T^{\mathrm{th}}=$140, 120 and 100 MeV, respectively.}
\label{fig:DNDPT}
\end{figure}
The transverse momentum spectra of negative charged hadrons observed by the STAR Collaboration \cite{STAR1} are compared with our results in Fig.~\ref{fig:DNDPT}.
It is found that the $p_t$ slope is insensitive to the thermal freeze-out temperature $T^{\mathrm{th}}$ for the model PCE while the $p_t$ slope from the model CE depends on $T^{\mathrm{th}}$.
This is a consequence of the suppression of radial flow due to the early chemical freeze-out \cite{Hirano1}.
We reproduce the $p_t$ spectra below 1.0-1.5 GeV/$c$ and underestimate in higher $p_t$ region in the model PCE analysis.
This indicates that the hard contribution becomes important above 1.5-2.0 GeV/$c$.
It should be noted that contribution from non-thermalized high $p_t$ partons with an appropriate amount of energy loss can fill in the difference between hydro results and the experimental data in the moderate high $p_t$ region ($2 < p_t < 5 $ GeV/$c$) \cite{Hirano3}. 
\begin{figure}[tb]
\begin{center}
\includegraphics[width=70mm]{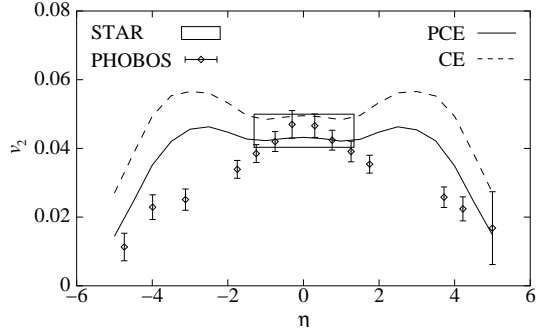}
\end{center}
\caption{$v_2(\eta)$ of charged hadrons in Au + Au central collisions at $\sqrt{s_{NN}}=130$ GeV for the model CE (dashed) and for the model PCE (solid).}
\label{fig:V2H}
\end{figure}
We next show in Fig.~\ref{fig:V2H} the pseudorapidity dependence of the second Fourier coefficient of azimuthal distribution $v_2(\eta)$.
Here we choose $T^{\mathrm{th}}=140$ MeV for both models. Experimental data by PHOBOS \cite{PHOBOS2} and STAR \cite{STAR2} are compared with our results.
We see, as well as the radial flow, the elliptic flow is suppressed by taking into account the early chemical freeze-out.
The shape of $v_2(\eta)$ is sensitive to the initial parameterization of energy density in the longitudinal direction.
Bumps in forward and backward rapidity regions result from the crescent shape of initial energy density in transverse plane at space-time rapidity $\mid \eta_{\mathrm{s}} \mid \sim$ 3-4 \cite{Hirano2}.
We might check whether other initial parameters can lead us to reproduce $dN/d\eta$ and $v_2(\eta)$ simultaneously.
\begin{figure}[tb]
\begin{minipage}[t]{75mm}
\includegraphics[width=70mm]{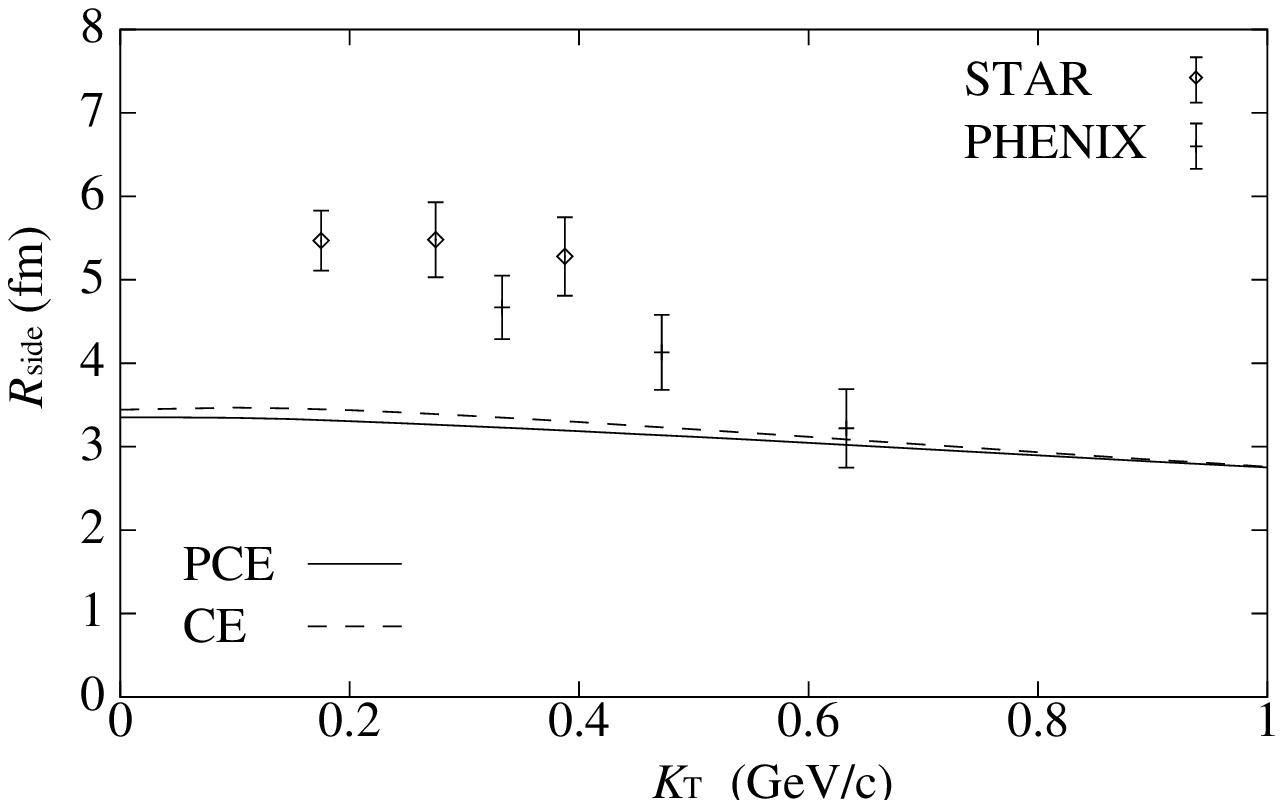}
\includegraphics[width=70mm]{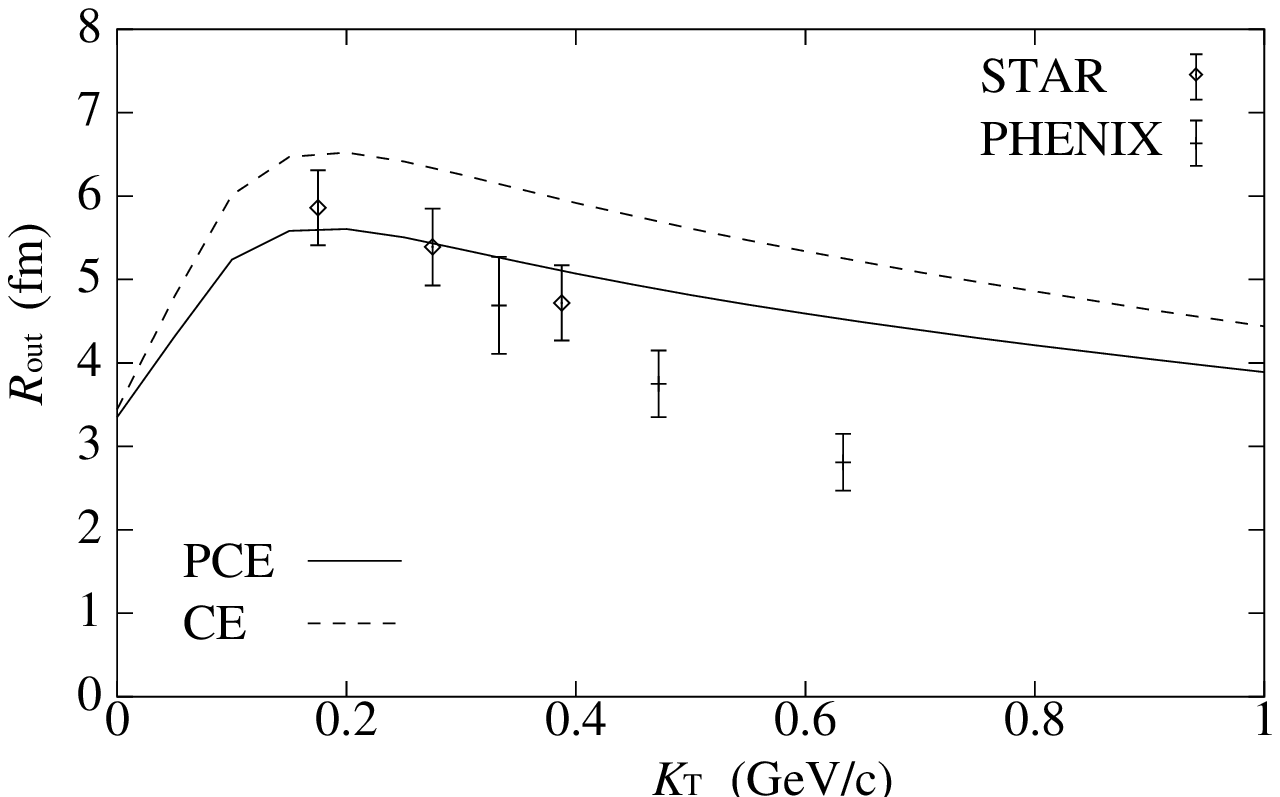}
\end{minipage}
\hspace{\fill}
\begin{minipage}[t]{75mm}
\includegraphics[width=70mm]{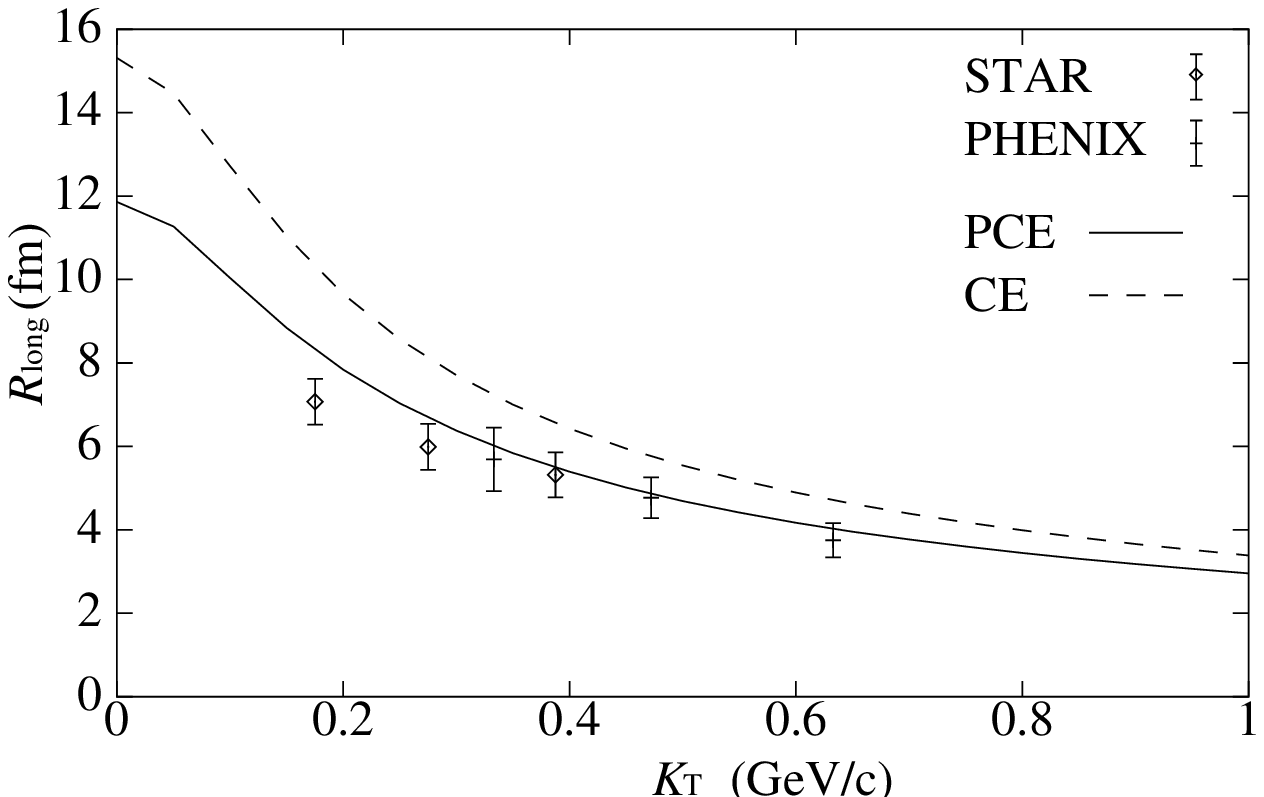}
\includegraphics[width=70mm]{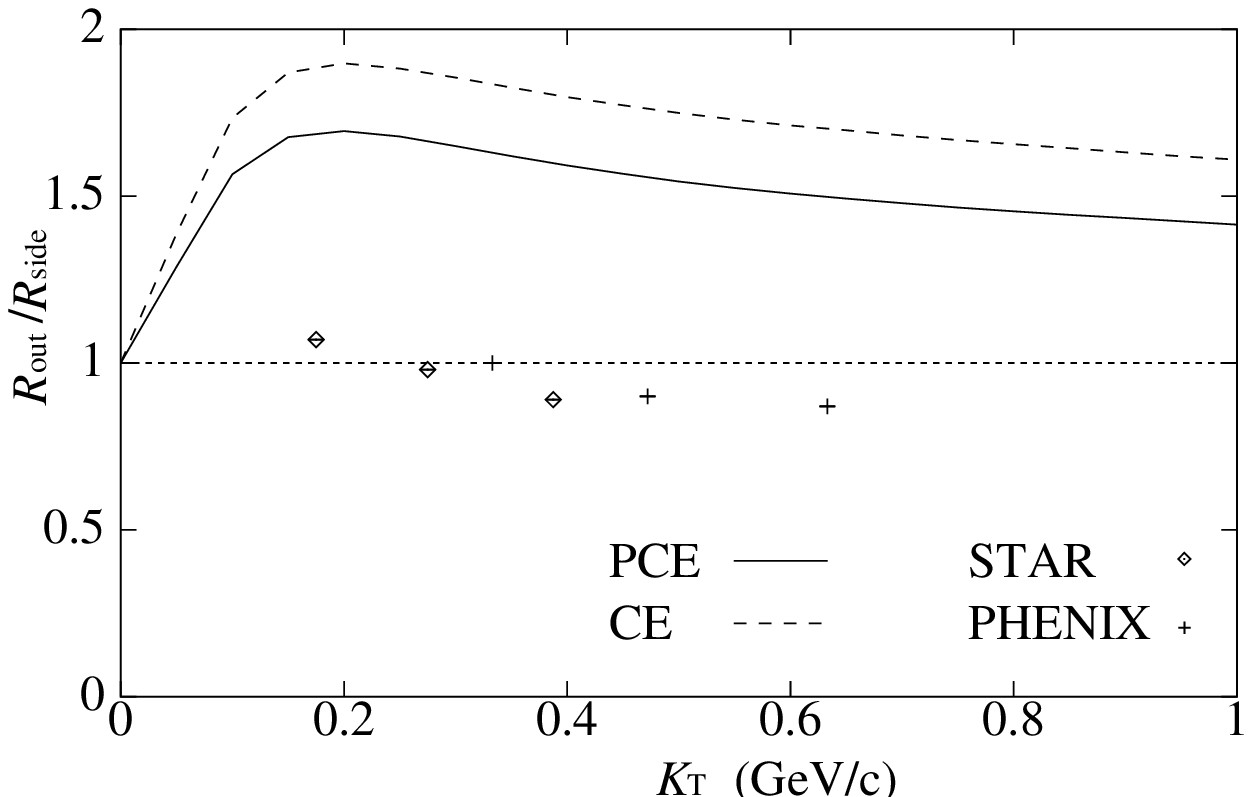}
\end{minipage}
\caption{HBT radii in Au + Au central collisions at $\sqrt{s_{NN}}=130$ GeV. The solid (dashed) lines correspond to the results from the model PCE (CE).}
\label{fig:HBT}
\end{figure}
Finally, let us see the effect of the early chemical freeze-out on the HBT radii. Figure \ref{fig:HBT} shows $R_{\mathrm{side}}$, $R_{\mathrm{out}}$, $R_{\mathrm{long}}$ and $R_{\mathrm{out}}/R_{\mathrm{side}}$ as functions of the transverse momentum of negative pion pair $K_T$.
We compare our results with experimental data by STAR \cite{STAR3} and PHENIX \cite{PHENIX}.
$R_{\mathrm{out}}$ and $R_{\mathrm{long}}$ are reduced by the early chemical freeze-out.
This mainly comes from the reduction of the lifetime of a fluid \cite{Hirano1}.
We reproduce the $K_T$ dependence of $R_{\mathrm{long}}$ by the model PCE, so the longitudinal dynamics can be correctly described by a hydrodynamic model with the finite longitudinal size and the chemical non-equilibrium property.
On the other hand, we cannot reproduce the transverse radii even in this model.
Therefore the early chemical freeze-out is not the complete interpretation of the RHIC HBT puzzle.

\section{SUMMARY}
We investigated the effect of early chemical freeze-out on collective flow and HBT radii at the RHIC energy with a full 3D hydrodynamic model.
It is found that radial flow, elliptic flow, $R_{\mathrm{out}}$ and $R_{\mathrm{long}}$ are reduced by considering the early chemical freeze-out.
In this analysis, we took $\mu_{\mathrm{B}} = 0$ as a first trial calculation.
It might be interesting to construct a model PCE with $\mu_{\mathrm{B}} \neq 0$ and to discuss particle ratios and spectra simultaneously at various collision energies. This will be discussed elsewhere.

\end{document}